\documentclass[prd,showpacs,nofootinbib,preprintnumbers]{revtex4}
\usepackage{amsmath}
\usepackage{amsfonts}
\usepackage{graphicx}
\usepackage{dcolumn}
\usepackage{hyperref}

\def\be{\begin{equation}}
\def\ee{\end{equation}}
\def\bea{\begin{eqnarray}}
\def\eea{\end{eqnarray}}

\def\5{\overline 5}

\begin{document}

\title{Statefinder diagnosis for the Palatini $f(R)$ gravity theories}

\author{Song Li$^{\,1}$, Hao-Ran Yu$^{\,2}$ and Tong-Jie Zhang$^{\,2,3,4,}$\footnote{\,Email address: tjzhang@bnu.edu.cn}}
\affiliation{$^1$Department of Physics, Beijing Normal University,
Beijing 100875, P. R. China\\
$^2$Department of Astronomy, Beijing Normal University,
Beijing 100875, P. R. China\\
$^3$Center for High Energy Physics, Peking University, Beijing
100871,
P. R. China\\
$^4$Kavli Institute for theoretical Physics China, Institute of
Theoretical
 Physics,
Chinese\\ Academy of Sciences (KITPC/ITP-CAS), P. O. Box 2735,
Beijing 100080, P. R. China}

\begin{abstract}
The Palatini $f(R)$ gravity, is able to probably explain the late
time cosmic acceleration without the need for dark energy, is
studied. In this paper, we investigate a number of $f(R)$ gravity
theories in Palatini formalism by means of statefinder diagnosis. We
consider two types of $f(R)$ theories: (i) $f(R)=R+\alpha
R^{m}-\beta R^{-n}$ and (ii) $f(R)=R+\alpha ln R+\beta$. We find
that the evolutionary trajectories in the $s-r$ and $q-r$ planes for
various types of the Palatini $f(R)$ theories reveal different
evolutionary properties of the universe. Additionally, we use the
observational $H(z)$ data to constrain models of $f(R)$ gravity.

\end{abstract}

\maketitle

\section{Introduction}
Recently, the discovery of the acceleration of cosmological
expansion at present epoch has been the most principal achievement
of observational cosmology. Numerous cosmological observations, such
as Type Ia Supernovae (SNIa)~\cite{price1}, Cosmic Microwave
Background (CMB)~\cite{price2} and Large Scale
Structure~\cite{price3}, strongly suggest that the universe is
spatially flat with about $4\%$ ordinary baryonic matter, $20\%$
dark matter and $76\%$ dark energy. The accelerated expansion of the
present universe is attributed to the dominant component of the
universe, dark energy, which not only has a large negative pressure,
but also does not cluster as ordinary matter does. In fact, it has
not been detected directly and there is no justification for
assuming that dark energy resembles known forms of matter or energy.
A large body of recent work has focussed on understanding the nature
of dark energy. However, the physical origin of dark energy as well
as its nature remain enigmatic at present.

The simplest model of dark energy is the cosmological constant
$\Lambda$~\cite{price4}, whose energy density remains constant with
time $\rho_{\Lambda}=\Lambda/8\pi G$ (natural units $c=\hbar=1$ is
used throughout the paper) and whose equation of state (defined as
the ratio of pressure to energy density) remains $w=-1$ as the
universe evolves. Unfortunately, the model is burdened with the well
known cosmological constant problems, namely the fine-tuning
problem: why is the energy of the vacuum so much smaller than we
estimate it should be? and the cosmic coincidence problem: why is
the dark energy density approximately equal to the matter density
today? These problems has led many researchers to try a different
approach to the dark energy issue. Furthermore, the recent analysis
of the SNIa data indicates that the time dependent dark energy gives
better fit than the cosmological constant. Instead of assuming the
equation of state $w$ is a constant, some authors investigate the
dynamical scenario of dark energy which is usually described by the
dynamics of a scalar field. The most popular model among them is
dubbed quintessence~\cite{price5}, which invoke an evolving scalar
field $\phi$ with a self-interaction potential $V(\phi)$ minimally
coupled to gravity. Besides, other scalar-field dark energy models
have been studied, including phantom~\cite{price6},
tachyon~\cite{price7}, quintom~\cite{price8}, ghost
condensates~\cite{price9}, etc. Also, there are other candidates,
for example, Chaplygin gas which attempt to unify dark energy and
dark matter~\cite{price10}, braneworld model which explain the
acceleration through the fact that the general relativity is
formulated in five dimensions instead of the usual
four~\cite{price11}.

On the other hand, recently, more and more researchers have made a
great deal of effort to consider modifying Einstein's general
relativity (GR) in order to interpret an accelerated expansion of
the universe avoiding the existence of dark energy. As is well
known, there are numerous ways to generalize Einstein's theory, in
which the most famous alternative to GR is scalar-tensor
theory~\cite{price12}. There are still various proposals, for
example, Dvali-Gabadadze-Porrati (DGP) gravity~\cite{price13},
$f(R)$ gravity~\cite{price14}, and so forth. The so-called $f(R)$
gravity is a straightforward generalization of the Einstein-Hilbert
action by including nonlinear terms in the scalar curvature. It has
been shown that some of these additional terms can give accelerating
expansion without dark energy~\cite{price15}.

Generally, in deriving the Einstein field equations there are
actually two different variational principles that one can apply to
the Einstein-Hilbert action, namely, the metric and the Palatini
approach. The choice of the variational principle is usually
referred to as a formalism, so one can use the metric formalism and
the Palatini formalism. In the metric formalism, the connection is
assumed to be the Christoffel symbol defined in terms of the metric
and the action is only varied with respect to the metric. While in
the latter the metric and the connection are treated as independent
variables and one varies the action with respect to both of them. In
fact, for an action which is linear in $R$, both approaches are
equivalent, and the theory reduces to GR. However when the action
includes nonlinear functions of of the Ricci scalar $R$, the two
methods give different field equations.

It was pointed out by Dolgov and Kawasaki that the fourth order
equations in the metric formalism suffer serious instability
problem~\cite{price16}, however, the Palatini formalism provides
second order field equations, which are free from the instability
problem mentioned above~\cite{price17}. Additionally, for the metric
approach, the models of the type $f(R)=R-\beta/R^{n}$ are
incompatible with the solar system experiments~\cite{price18} and
have the correct Newtonian limit seemed to be a controversial
issue~\cite{price19}. Another important point is that these models
can not produce a standard matter-dominated era followed by an
accelerating expansion~\cite{price20,price21}. While, for the
Palatini approach the models satisfy the solar system tests but also
have the correct Newtonian limit~\cite{price22}. Furthermore, in
Ref.~\cite{price23} it has been shown that the above type can
produce the sequence of radiation-dominated, matter-dominated and
late accelerating phases. Thus, as already mentioned, the Palatini
approach seems appealing though some issues are still of debate, for
example, the instability problems~\cite{price22,price24}. Here, we
will concentrate on the Palatini formalism.

In addition, since more and more cosmological models have been
proposed, the problem of discriminating different dark energy models
is now emergent. In order to solve this problem, a sensitive and
robust diagnosis for dark energy models is necessary. As we all
know, the equation of state $w$ could probably discriminate some
basic dark energy models, i.e., the cosmological constant $\Lambda$
with $w=-1$, the quintessence with $w>-1$, the phantom with $w<-1$,
and so on. However, for some geometrical models arising from
modifications to the gravitational sector of Einstein's theory, the
equation of state $w$ no longer plays the role of a fundamental
physical quantity and the ambit of it is not so clear, thus it would
be very useful to propose a new diagnosis to give all classes of
cosmological models an unambiguous discrimination. In order to
achieve this aim, Sahni et al.~\cite{price25} introduced the
statefinder pair $\{r,s\}$, where $r$ is generated from the scalar
factor $a$ and its higher derivatives with respect to the cosmic
time $t$, and $s$ is expressed by $r$ and the deceleration parameter
$q\equiv-a\ddot{a}/\dot{a}^{2}$. Therefore, the statefinder is a
``geometrical'' diagnostic in the sense that it depends upon the
scalar factor and hence upon the metric describing space time.
According to different cosmological models, clear differences for
the evolutionary trajectories in the $s-r$ plane can be found, so
the statefinder diagnostic may possibly be used to discriminate
different cosmological models. In recent works~\cite{price26}, the
statefinder diagnostic have been successfully demonstrated that it
can differentiate a series of cosmological models ,including the
cosmological constant, the quintessence, the phantom, the Chaplygin
gas, the holographic dark energy models, the interacting dark energy
models, etc.

In this paper, we focus on the $f(R)$ theory in Palatini formalism
and consider a number of varieties of $f(R)$ theories recently
proposed in the literature. Moreover, we apply the statefinder
diagnosis to such $f(R)$ theories. We find that the models in the
Palatini $f(R)$ gravity can be distinguished from dark energy
models. In addition, we use the observational $H(z)$ data derived
from ages of the passively evolving galaxies to make a combinational
constraint.

This paper is organized as follows: In Sec.2, we briefly review the
$f(R)$ gravity in Palatini formalism and study the cosmological
dynamical behavior of Palatini $f(R)$ theories. In Sec.3, we apply
the statefinder diagnosis to various $f(R)$ gravity. In Sec.4, we
obtain the parameters from the observational constraints. Finally,
the conclusions and the discussions are presented.

\section{The Palatini $f(R)$ Gravity and Its Cosmology}

\subsection{a brief overview of $f(R)$ gravity in Palatini formalism}
We firstly review the Palatini formalism from the generalized
Einstein-Hilbert action
\begin{equation}
S=\frac{1}{2\kappa}\int d^{4}x\sqrt{-g}f(R)+S_{m}(g_{\mu\nu},\psi),
\end{equation}
where $\kappa\equiv 8\pi G$, $G$ is the gravitational constant, $g$
is the determinant of the metric $g_{\mu\nu}$, $f(R)$ is the general
function of the generalized Ricci scalar $R\equiv
g^{\mu\nu}R_{\mu\nu}({\Gamma^{\lambda}}_{\mu\nu})$ and $S_{m}$ is
the matter action which depends only on the metric $g_{\mu\nu}$ and
the matter fields $\psi$ and not on the independent connection
${\Gamma^{\lambda}}_{\mu\nu}$ differentiated from the Levi-Civita
connection $\{{^{\lambda}}_{\mu\nu}\}$. In our paper, nature unit
$c=1$ is used. It should be noted that GR will come about when
$f(R)=R$.

Varying the action with respect to the metric $g_{\mu\nu}$ and the
connection ${\Gamma^{\lambda}}_{\mu\nu}$ respectively yields
\begin{eqnarray}
&&f'(R)R_{\mu\nu}-\frac{1}{2}f(R)g_{\mu\nu}=\kappa T_{\mu\nu},\\
&&\overline{\nabla}_{\lambda}(\sqrt{-g}f'(R)g^{\mu\nu})=0,
\end{eqnarray}
where $f'(R)\equiv df/dR$, $\overline{\nabla}_{\lambda}$ denotes the
covariant derivative associated with the independent connection
${\Gamma^{\lambda}}_{\mu\nu}$ and $T_{\mu\nu}$ is the
energy-momentum tensor given by
\begin{equation}
T_{\mu\nu}=-\frac{2}{\sqrt{-g}}\frac{\delta S_{m}}{\delta
g^{\mu\nu}}.
\end{equation}
If we consider a perfect fluid, then
$T^{\mu\nu}=(\rho+p)u^{\mu}u^{\nu}+pg^{\mu\nu}$, where $\rho$ and
$p$ respectively denotes the energy density and the pressure of the
fluid, $u^{\mu}$ is the fluid four-velocity. Note that
$T=g^{\mu\nu}T_{\mu\nu}=-\rho+3p$.

According to Eq. (3), we can define a metric conformal to
$g_{\mu\nu}$ as
\begin{equation}
h_{\mu\nu}\equiv f'(R)g_{\mu\nu}.
\end{equation}
Then, we can get the connection ${\Gamma^{\lambda}}_{\mu\nu}$ in
terms of the conformal metric $h_{\mu\nu}$
\begin{equation}
{\Gamma^{\lambda}}_{\mu\nu}=h^{\lambda\sigma}(h_{\sigma\mu,\nu}+h_{\sigma\nu,\mu}-h_{\mu\nu,\sigma}),
\end{equation}
furthermore, it can be equally written as
\begin{equation}
{\Gamma^{\lambda}}_{\mu\nu}=\frac{1}{f'(R)}g^{\lambda\sigma}[\partial_{\mu}(f'(R)g_{\nu\sigma})+\partial_{\nu}(f'(R)g_{\mu\sigma})-\partial_{\sigma}(f'(R)g_{\mu\nu})].
\end{equation}
Meanwhile, the generalized Ricci tensor is
\begin{equation}
R_{\mu\nu}={\Gamma^{\lambda}}_{\mu\nu,\lambda}-{\Gamma^{\lambda}}_{\mu\lambda,\nu}+{\Gamma^{\lambda}}_{\lambda\sigma}{\Gamma^{\sigma}}_{\mu\nu}-{\Gamma^{\lambda}}_{\mu\sigma}{\Gamma^{\sigma}}_{\lambda\nu},
\end{equation}
and thus, we can rewrite the generalized Ricci tensor expressed by
the Ricci tensor $R_{\mu\nu}(g)$ associated with $g_{\mu\nu}$ as
\begin{eqnarray}
R_{\mu\nu}&=&R_{\mu\nu}(g)+\frac{3}{2}\frac{1}{f'(R)^{2}}(\nabla_{\mu}f'(R))(\nabla_{\nu}f'(R))\nonumber\\
&&-\frac{1}{f'(R)}\nabla_{\mu}\nabla_{\nu}f'(R)-\frac{1}{2}\frac{1}{f'(R)}g_{\mu\nu}\nabla_{\sigma}\nabla^{\sigma}f'(R),
\end{eqnarray}
where $\nabla_{\mu}$ is the covariant derivative defined with the
Levi-Civita connection of the metric.

\subsection{FRW cosmology of the Palatini $f(R)$ gravity and numerical results }
Since measurements of CMB suggest that our universe is spatially
flat~\cite{price27,price28}, we start our work with a flat
Friedmann-Robertson-Walker (FRW) universe with metric
\begin{equation}
ds^{2}=-dt^{2}+a^{2}(t)(dx^{2}+dy^{2}+dz^{2}),
\end{equation}
where $a(t)$ is the scalar factor and $t$ is the cosmic time.

As a result, making use of Eqs. (2) and (9), the modified Friedmann
equation can be derived as
\begin{equation}
(H+\frac{1}{2}\frac{\dot{f}'(R)}{f'(R)})^{2}=\frac{1}{6}\frac{\kappa(\rho+3p)+f(R)}{f'(R)},
\end{equation}
where $H$ is the Hubble parameter and the dot denotes the
differentiation with respect to the cosmic time $t$.

In addition, taking the trace of Eq. (2), gives
\begin{equation}
f'(R)R-2f(R)=\kappa T.
\end{equation}
If we consider the universe only containing dust-like matter, then
$T=-\rho_{m}$, where, $\rho_{m}$ denotes the energy density of
matter. Furthermore, combining Eq. (12) with the energy conservation
equation of matter $\dot{\rho}_{m}+3H\rho_{m}=0$, we can express
$\dot{R}$ as
\begin{equation}
\dot{R}=-\frac{3H(f'(R)R-2f(R))}{f''(R)R-f'(R)},
\end{equation}
where $f''(R)\equiv d^{2}f/dR^{2}$. Replacing $\dot{R}$ in Eq. (11)
with Eq. (13), yields
\begin{equation}
H^{2}=\frac{1}{6f'(R)}\frac{3f(R)-f'(R)R}{[1-\frac{3}{2}\frac{f''(R)(f'(R)R-2f(R))}{f'(R)(f''(R)R-f'(R))}]^{2}},
\end{equation}
from which, we can get the modified Friedmann equation with respect
to $R$ given the form of $f(R)$.

Using the redshift $z=\frac{1}{a}-1$ (usually, the scalar factor $a$
of today is defined $a_{0}=1$), the expression
$\rho_{m}=\rho_{m0}(1+z)^{3}$ and $\frac{dz}{dt}=-H(1+z)$, we can
rewrite Eqs. (12) and (13) as
\begin{eqnarray}
&&f'(R)R-2f(R)=-3H_{0}^{2}\Omega_{m0}(1+z)^{3},\\
&&\frac{dR}{dz}=-\frac{9H_{0}^{2}\Omega_{m0}(1+z)^{2}}{f''(R)R-f'(R)},
\end{eqnarray}
where $\Omega_{m0}\equiv\frac{\kappa\rho_{m0}}{3H_{0}^{2}}$ and the
subscript $0$ throughout the paper denotes the present time.
Therefore, we can express the Hubble parameter $H$ in terms of $z$
as
\begin{equation}
\frac{H^{2}}{H_{0}^{2}}=\frac{1}{6f'(R)}\frac{3\Omega_{m0}(1+z)^{3}+f(R)/H_{0}^{2}}{[1+\frac{9}{2}\frac{H_{0}^{2}\Omega_{m0}(1+z)^{3}f''(R)}{(f''(R)R-f'(R))f'(R)}]^{2}},
\end{equation}
if the form of $f(R)$ with respect to $R$ is given. In order to
study the cosmological evolution by using Eqs. (15)-(17), it is
necessary to give the initial conditions: ($R_{0}$, $H_{0}$,
$\Omega_{m0}$). From Eqs. (15) and (17), choosing units so that
$H_{0}=1$~\cite{price29}, we can solve for $R_{0}$ given that the
rest ones except one parameter are fixed.

On the other hand, in order to understand the cosmological evolution
behavior, it is useful to define the effective equation of state
\begin{equation}
w_{eff}=-1+\frac{2}{3}(1+z)\frac{H^{'}}{H}.
\end{equation}
Thus, we can show the effective equation of state as a function of
redshift $z$ for any $f(R)$ from Eqs. (16)-(18).

In our paper, we adopt two types of $f(R)$ theories, recently
considered in the literature.

\subsubsection{$f(R)$ theories with power-law term}
We consider the following general form for $f(R)$
\begin{equation}
f(R)=R+\alpha R^{m}-\beta R^{-n},
\end{equation}
where $m$ and $n$ are real constants with the same sign. Such
theories have been considered with the hope of explaining the early
and the late accelerated expansion of our universe
~\cite{price30,price31}. Note that not all combinations of $m$ and
$n$ are agreement with a flat universe with the early matter
dominated era followed by an accelerated expansion at late times. At
the early times of matter dominated, the universe is better
described by GR in order to avoid confliction with early-time
physics such as Big Bang Nucleonsynthesis (SSN) and CMB. This
implies that the modified Lagrangian should recover the standard GR
Lagrangian for large $R$, and hence we demand that $m<1$ and $n>-1$.
Now we use the values of $\alpha$ and $\beta$ to classify such
theories.

\textbf{Case 1 $\alpha\neq0,\beta=0$ }

In this case, (19) reduces to
\begin{equation}
f(R)=R-\beta R^{-n}.
\end{equation}
Substituting the form of (20) into Eqs. (16)-(18), with the present
fractional matter density $\Omega_{m0}=0.27$, the changing of the
Ricci curvature $R$ and the effective equation of state $w_{eff}$
with the redshift $z$ are plotted in Fig.1. It is to be noted that
the special case of $(\alpha,n)=(-4.38,0)$ corresponds to the
$\Lambda$CDM model. We can easily see that the curvature and the
effective equation of state decrease with the evolution of the
universe for any choice of $n$. Moreover, the smaller $n$, the
faster $R$ decrease, and the larger the present value of $w_{eff}$.
Also, the universe turns to an accelerated phase from a decelerated
era, and tends to a de Sitter phase in the future.

\begin{figure}[tbp]
\includegraphics[width=0.4\textwidth]{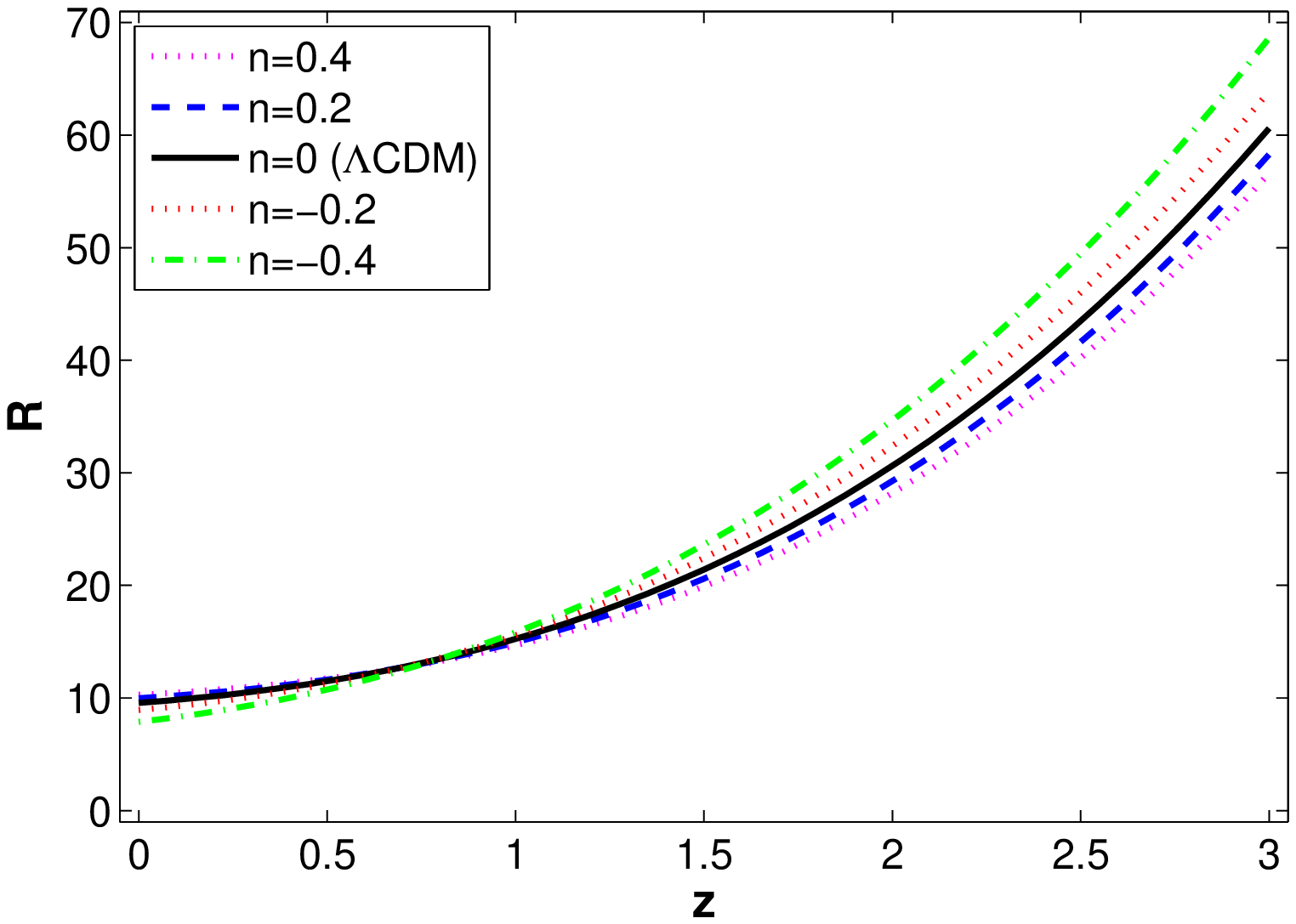}%
\includegraphics[width=0.4\textwidth]{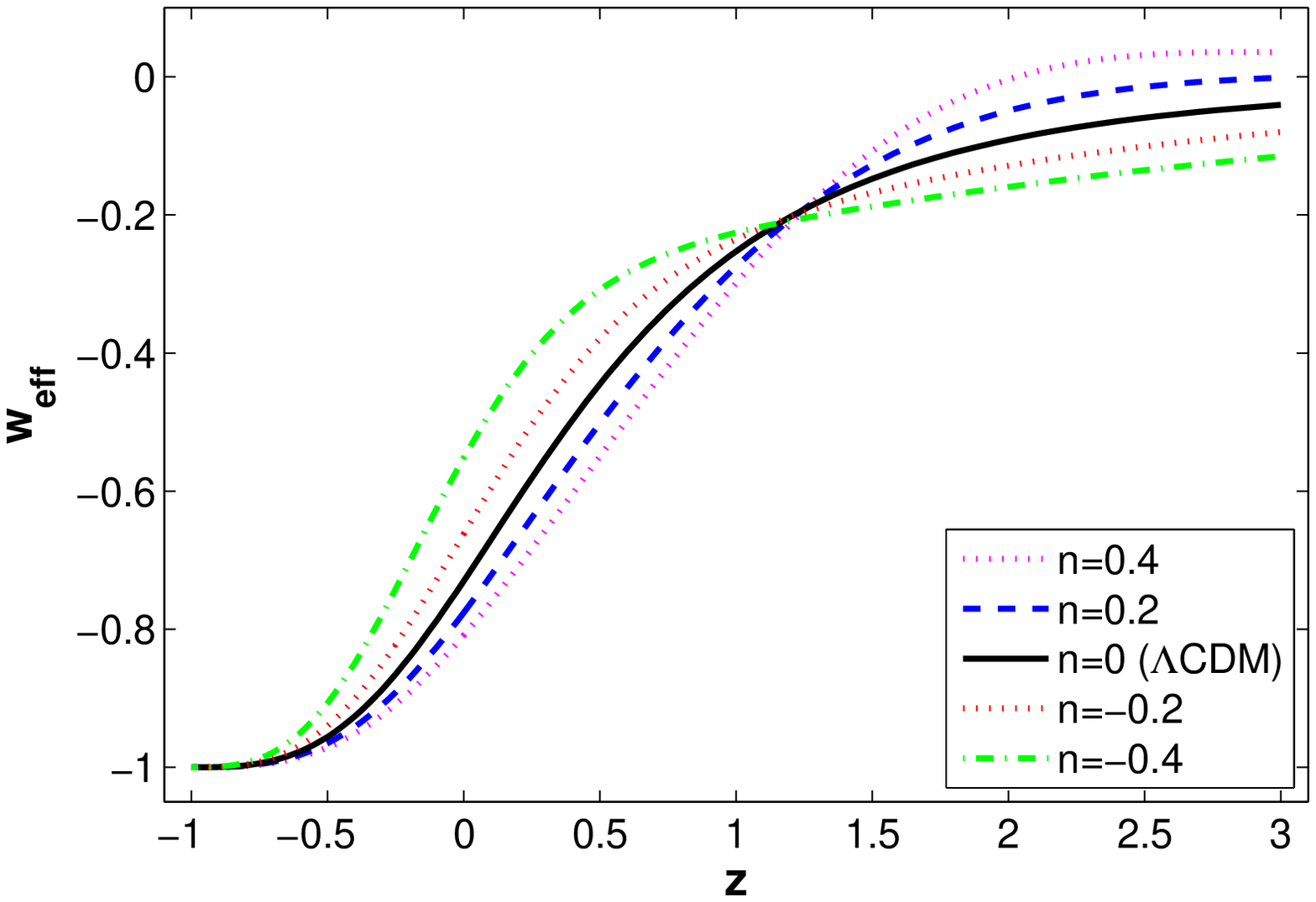}
\caption{The figures are for the model $f(R)=R-\beta R^{-n}$ .The
left figure is the diagram of Ricci scalar $R$ as a function of
redshift $z$. The right figure is the evolution trajectories of
$w_{eff}$. Different values of $n$ are selected as
$0.4,0.2,0,-0.2,-0.4$ with $\Omega_{m0}=0.27$. } \label{Fig.1}
\end{figure}

\begin{figure}[tbp]
\includegraphics[width=0.4\textwidth]{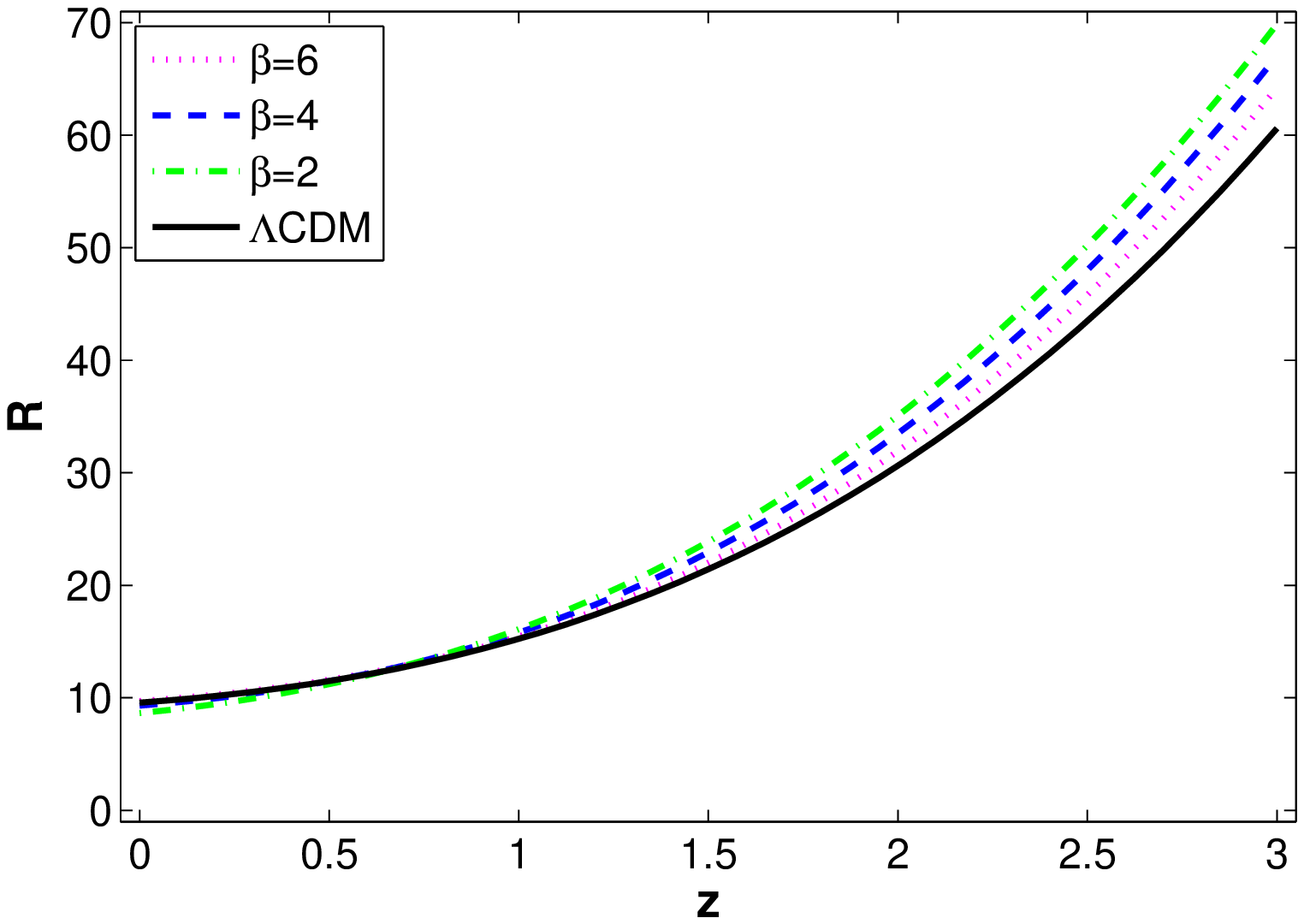}%
\includegraphics[width=0.4\textwidth]{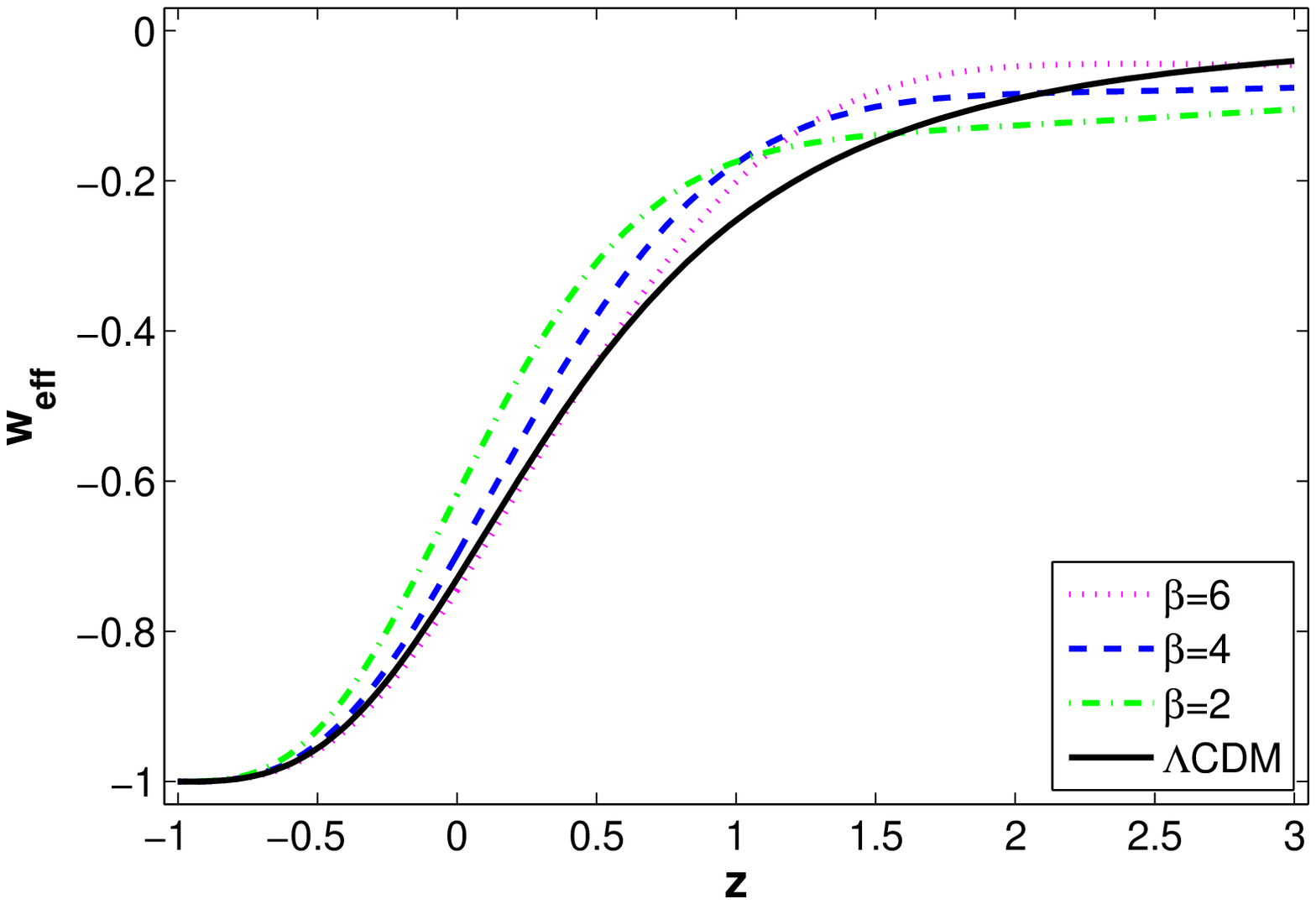}
\caption{The figures are for the model $f(R)=R+\alpha R^{m}-\beta
R^{-n}$ with $m=n=1/2$. The present fractional matter density
$\Omega_{m0}=0.27$. The left figure is the diagram of Ricci scalar
$R$ as a function of redshift $z$. The right figure is the evolution
trajectories of $w_{eff}$. } \label{Fig.2}
\end{figure}

\begin{figure}[tbp]
\includegraphics[width=0.4\textwidth]{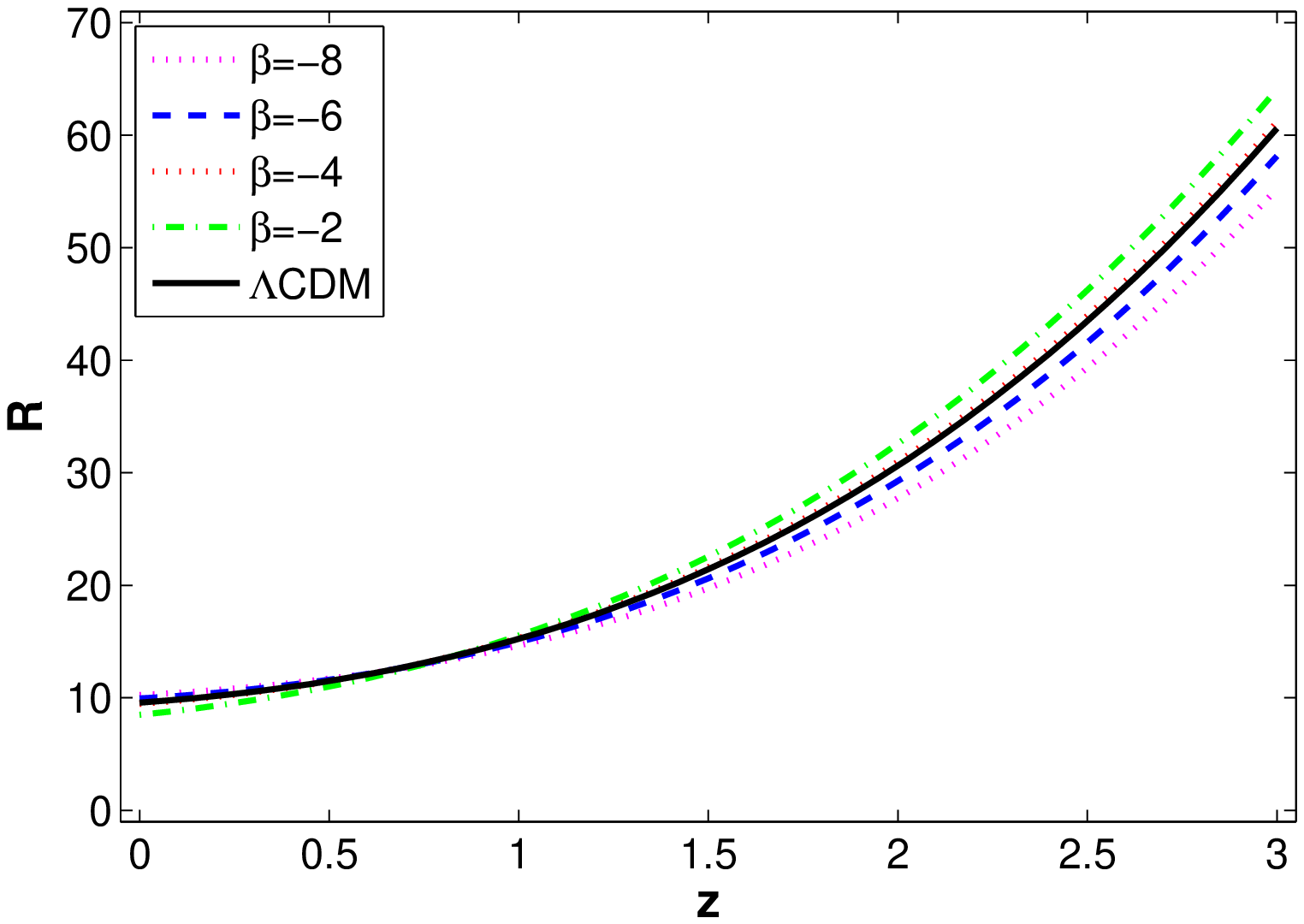}%
\includegraphics[width=0.4\textwidth]{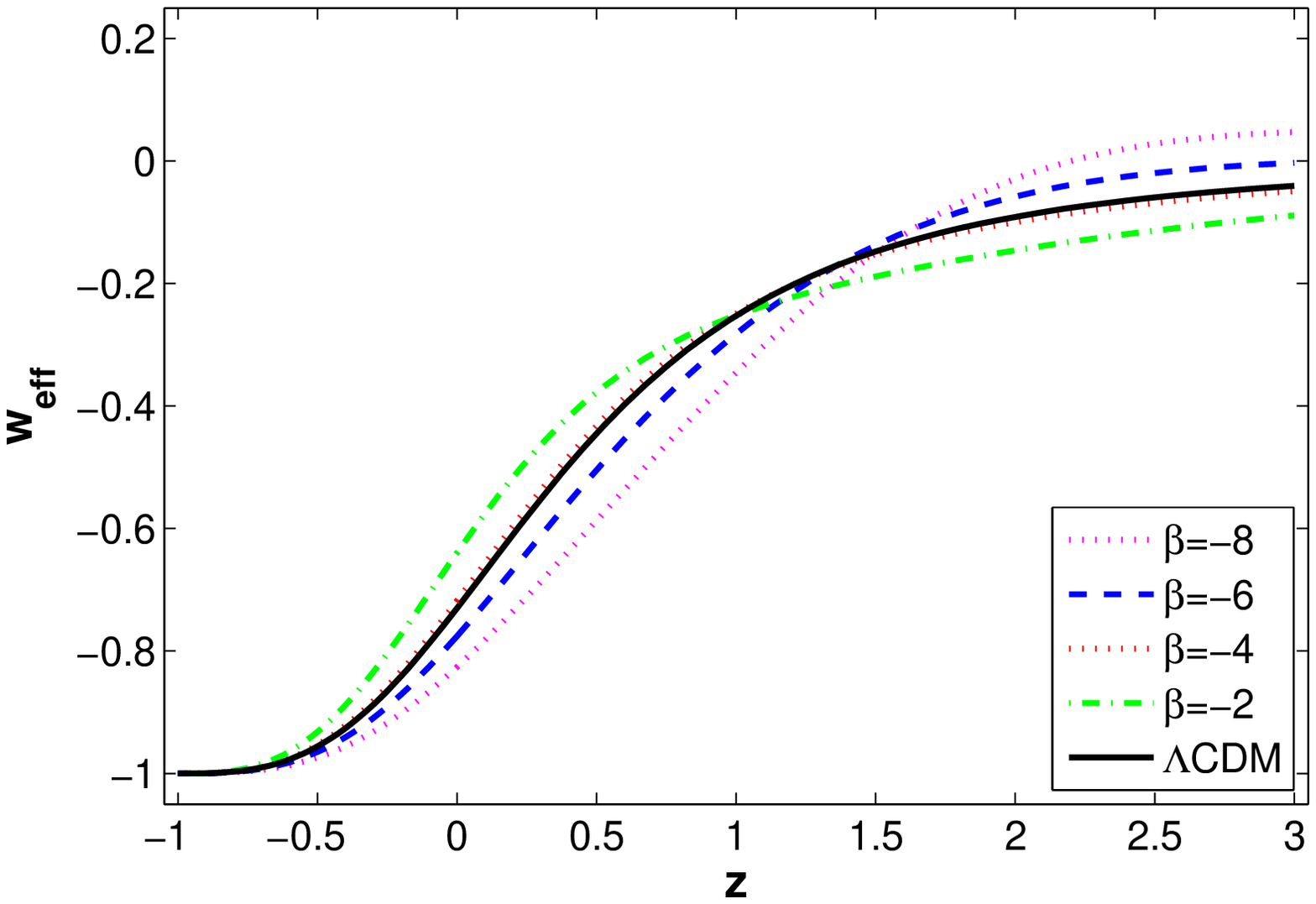}
\caption{The figures are for the model $f(R)=R+\alpha ln R+\beta$
.The left figure is the diagram of Ricci scalar $R$ as a function of
redshift $z$. The right figure is the evolution trajectories of
$w_{eff}$. } \label{Fig.3}
\end{figure}

\textbf{Case 2 $\alpha,\beta\neq0$}

In this case, we consider the general form of this type. To study
the cosmological behavior of such theories, we adopt the model
\begin{equation}
f(R)=R+\alpha R^{1/2}-\beta R^{-1/2}.
\end{equation}
From Fig.2, we clearly find that the curvature and the effective
equation of state decrease with the evolution of the universe for
any choice of $\beta$. the smaller $\beta$ results in the faster
decrease of $R$ and the larger $w_{eff}$ at the present time. It is
also obvious that the universe evolves from deceleration to
acceleration, and enters to a de Sitter phase in the future.

\subsubsection{$f(R)$ theories with logarithm term}
For this type of $f(R)$ theory, we adopt the form
\begin{equation}
f(R)=R+\alpha ln R+\beta,
\end{equation}
which has been studied in~\cite{price32,price33}. It has been
claimed that such theories have a well-defined Newtonian
limit~\cite{price32}. Note that, the asymptotic behavior $\lim_{R
\to +\infty}f(R) \to R$ is obtained for any choice of $\alpha$ and
$\beta$, and thus, the arbitrary $\alpha$ and $\beta$ can satisfy
the assumption that the universe is described by GR at the early
time. However, not all combinations of $\alpha$ and $\beta$ can
explain a late-time accelerated expansion of the universe.
Therefore, for the sake of compatibility with the observational
constraints obtained in Sec.4, we select a series values of $\beta$,
which can well explain the evolvement of the universe from an
early-time deceleration to a late-time acceleration (see Fig.3).

Substituting the form of (22) into Eqs. (16)-(18), with
$\Omega_{m0}=0.27$, the changing of the Ricci curvature $R$ and the
effective equation of state $w_{eff}$ with the redshift $z$ for this
model are plotted in Fig.3. Obviously, $R$ and $w_{eff}$ decrease
with the evolution of the universe for any choice of $\beta$. Also,
the larger $\beta$ results in the faster decrease of $R$ and the
larger $w_{eff}$ at the present time. It is noting that, similar to
the result of the above type of theories, the universe evolves from
deceleration to acceleration, and enters to a de Sitter phase in the
future.

\section{Statefinder Diagnosis for the Palatini $f(R)$ Gravity}
In this section, we turn our attention to the statefinder diagnosis.
As we know, two famous geometrical variables characterizing the
expansion history of the universe are the Hubble parameter
$H=\dot{a}/a$ describing the expansion rate of the universe and the
deceleration parameter $q=-\ddot{a}/aH^{2}$ characterizing the rate
of acceleration/deceleration of the expanding universe. It is clear
that they only depend on the scalar factor $a$ and its derivatives
with respect to $t$, i.e., $\dot{a}$ and $\ddot{a}$. However, as the
enhancing of cosmological models and the remarkable increase in the
accuracy of cosmological observational data, these variables are no
longer to be a perfect choice. This can be easily seen from the fact
that many cosmological models correspond to the same current value
of $q$. As a result, the so-called statefinder diagnosis was
introduced in order to discriminate more and more cosmological
models.

The statefinder diagnosis is constructed from the scalar $a$ and its
derivatives up to the third order. Namely, the statefinder pair
$\{r,s\}$ is defined as
\begin{equation}
r\equiv\frac{\dddot{a}}{aH^{3}},\mbox{}\hspace{15pt}s\equiv\frac{r-1}{3(q-1/2)}.
\end{equation}
Since different cosmological models exhibit qualitatively different
trajectories of evolution in the $s-r$ plane, the statefinder
diagnosis is a good tool to distinguish cosmological models. The
remarkable property is that the statefinder pair $\{r,s\}=(0,1)$
corresponds to the $\Lambda$CDM model. We can clearly identify the
``distance'' from a given cosmological model to $\Lambda$CDM model
in the $s-r$ plane, such as the quintessence, the phantom, the
Chaplygin gas, the holographic dark energy models, the interacting
dark energy models, and so forth, which have been shown in the
literatures~\cite{price26}. Particularly, the current values of the
parameters $s$ and $r$ in these diagrams can provide a consider way
to measure the ``distance'' from a given model to $\Lambda$CDM
model. Generally, according to the reexpression of the deceleration
parameter $q$
\begin{equation}
q=-1+(1+z)\frac{H'}{H},
\end{equation}
where $H'\equiv\frac{dH}{dz}$, we can also rewrite the statefinder
pair $\{r,s\}$ in terms of the Hubble parameter $H$ and its first
and second derivatives $H'$ and $H''$ with respect to the redshift
$z$ as
\begin{eqnarray}
r&=&1-2(1+z)\frac{H'}{H}+(1+z)^{2}{\frac{H'}{H}}^{2}+(1+z)^{2}\frac{H''}{H},\\
s&=&\frac{-2(1+z)H'/H+(1+z)^{2}{(H'/H)}^{2}+(1+z)^{2}H''/H}{3[(1+z)H'/H-3/2]}.
\end{eqnarray}

\begin{figure}[tbp]
\includegraphics[width=0.4\textwidth]{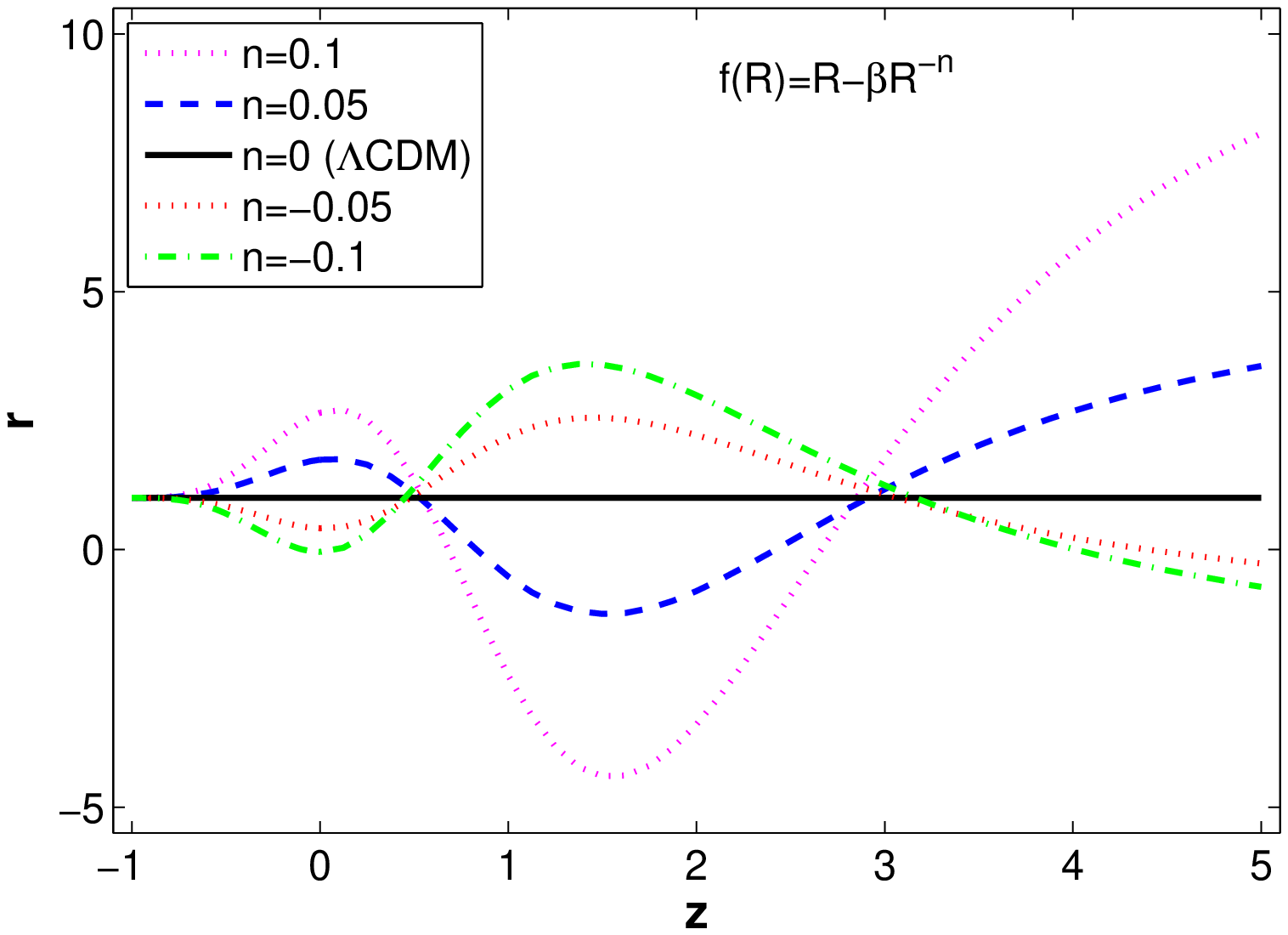}%
\includegraphics[width=0.4\textwidth]{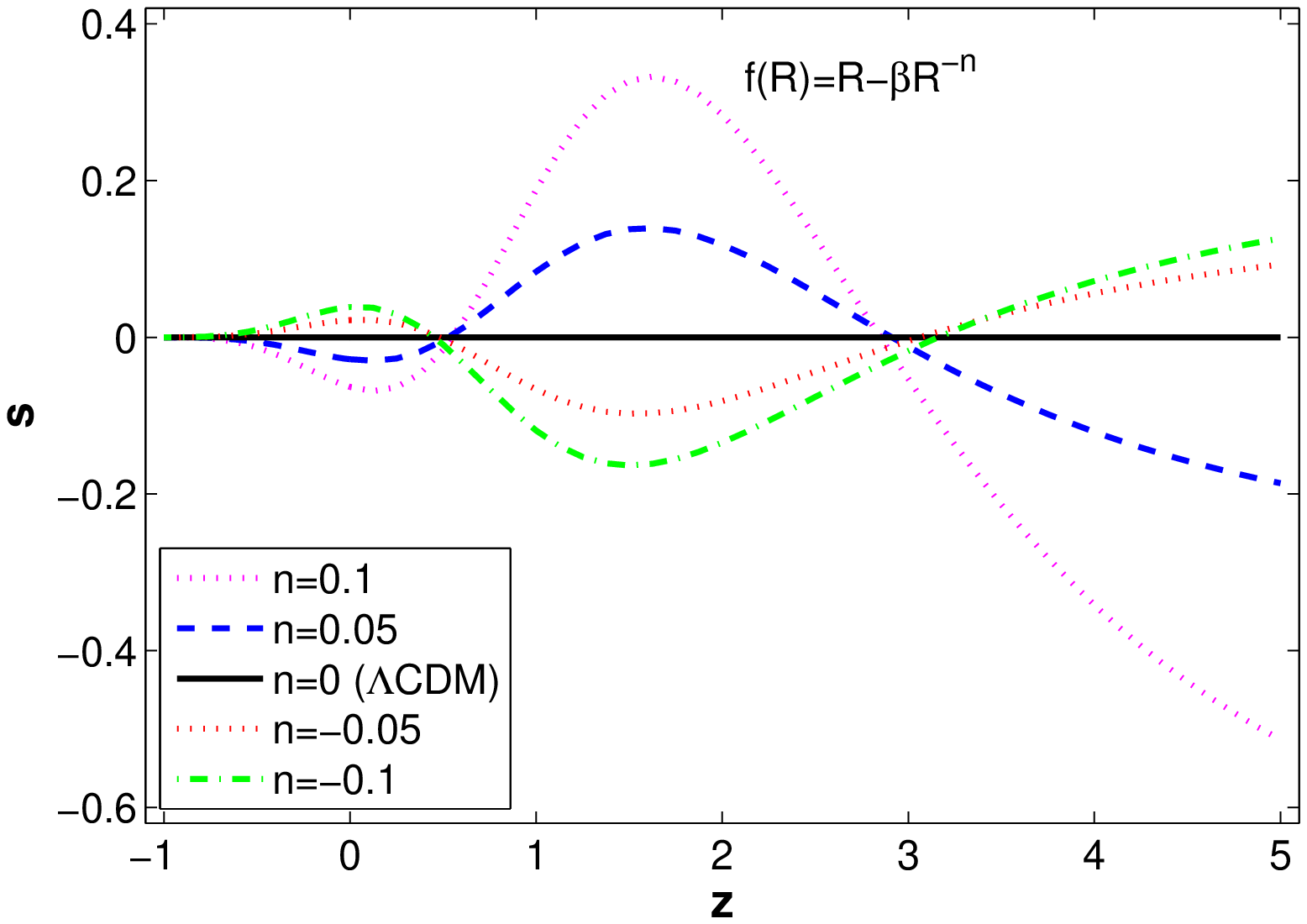}\\
\begin{center}
(a)
\end{center}
\includegraphics[width=0.4\textwidth]{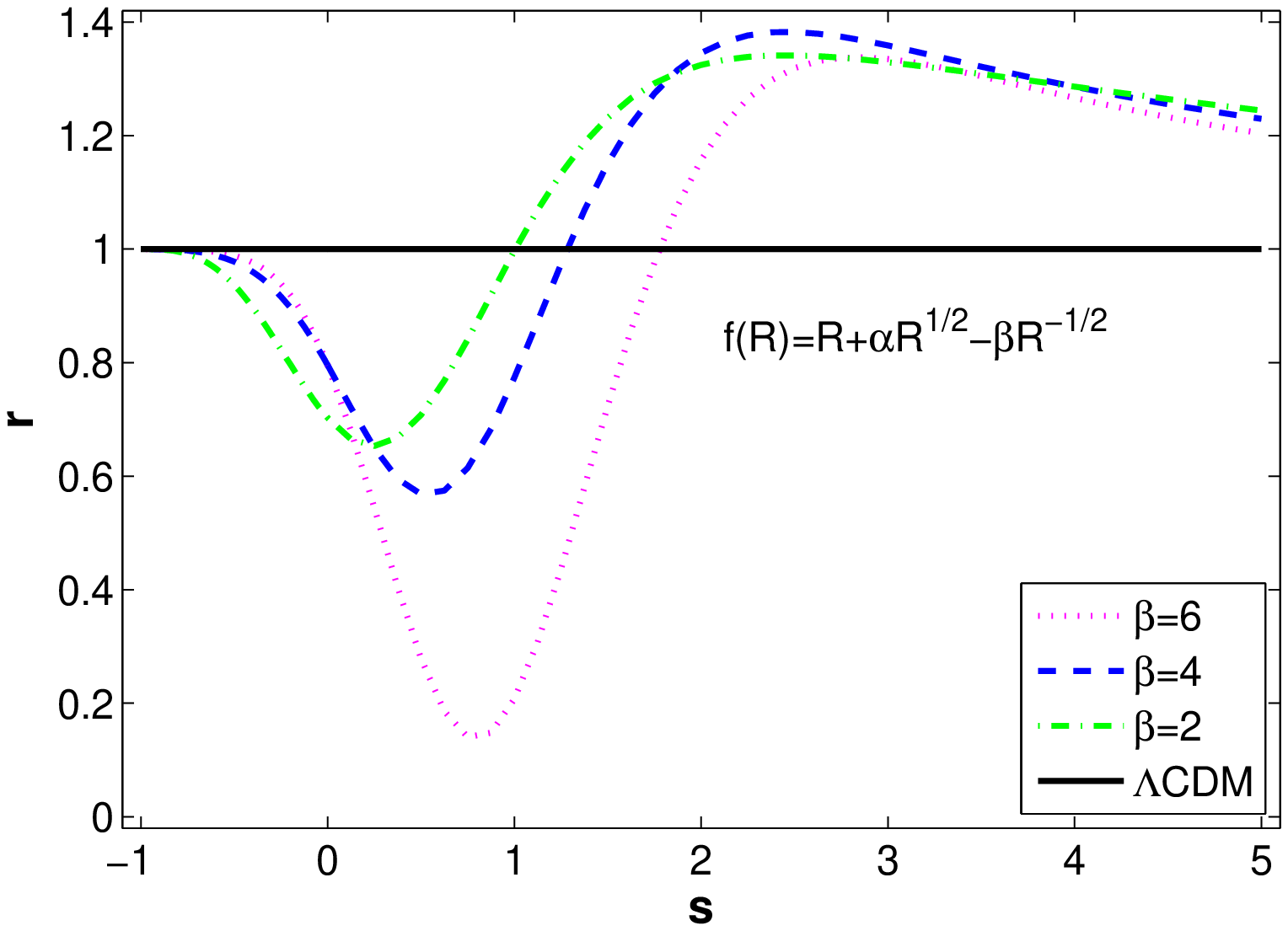}%
\includegraphics[width=0.4\textwidth]{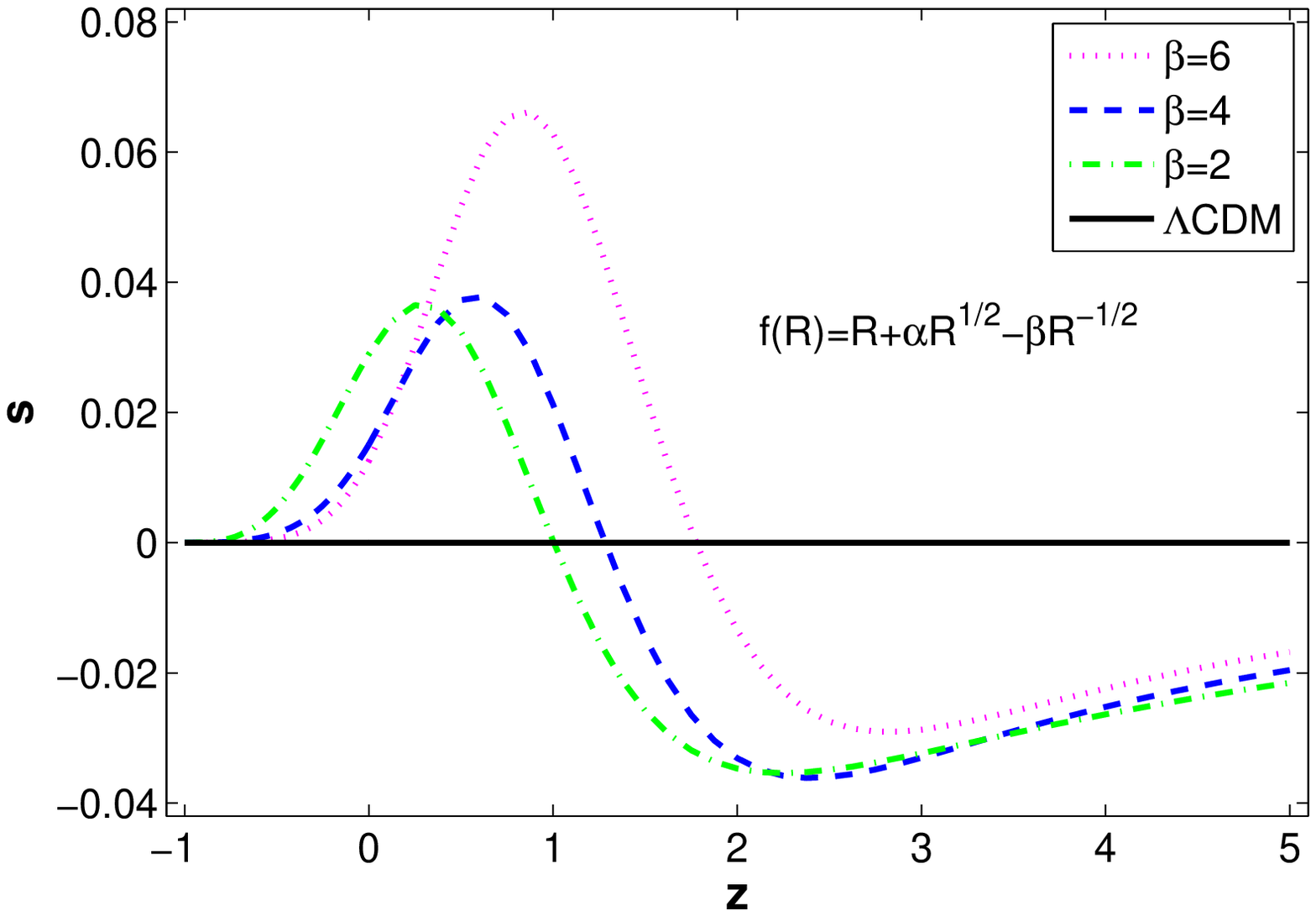}\\
\begin{center}
(b)
\end{center}
\includegraphics[width=0.4\textwidth]{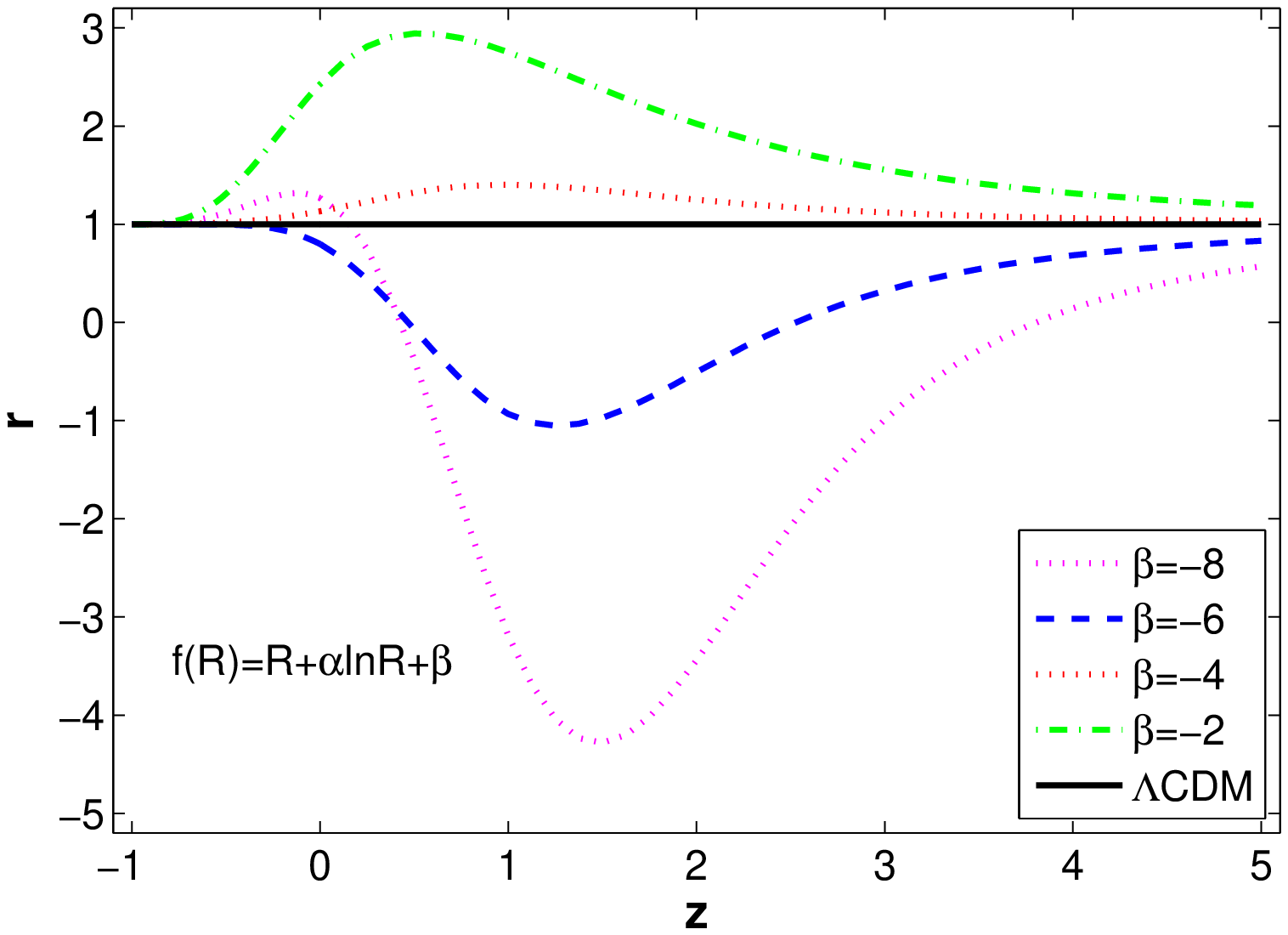}%
\includegraphics[width=0.4\textwidth]{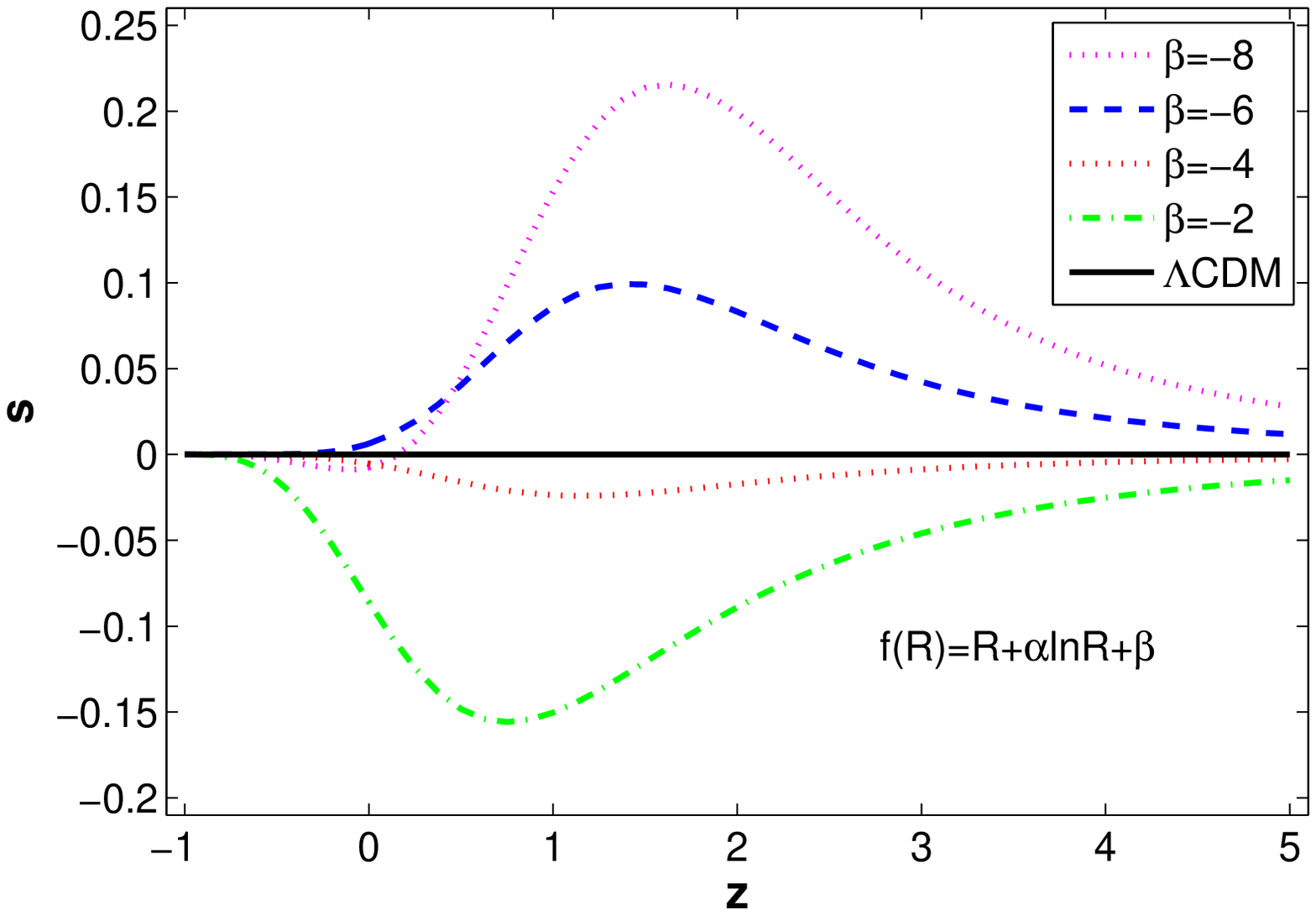}\\
\begin{center}
(c)
\end{center}
\caption{$r(z)$ and $s(z)$ for the models $f(R)=R-\beta R^{-n}$,
$f(R)=R+\alpha R^{1/2}-\beta R^{-1/2}$ and $f(R)=R+\alpha lnR+\beta$
with $\Omega_{m0}=0.27$. } \label{Fig.4}
\end{figure}

In what follows, we will apply the statefinder diagnosis to the
$f(R)$ theories mentioned in Sec.2. The values of parameters we
select in such theories are as same as those in the previous
section. By describing the evolution trajectories of statefinder
parameters $r$ and $s$ for the Palatini $f(R)$ theories, we can
differentiate $f(R)$ models from dark energy models, and even
discriminate various types of Palatini $f(R)$ theories from each
other.

We plot the statefinder parameters $r(z)$ and $s(z)$ for the above
two types of Palatini $f(R)$ theories in Fig.4 with
$\Omega_{m0}=0.27$. Fig.4 show that $r(z)$ and $s(z)$ for the three
models exhibit different features. For the model $f(R)=R-\beta
R^{-n}$ (see Fig.4a), the curves cross the $\Lambda$CDM line ($r=1$
or $s=0$) twice times and approach it in the future. Moreover, the
curves with $n>0$ start from the region $r>1,s<0$, while the other
way round, the curves with $n<0$ start from the region $r<1,s>0$.
For the model $f(R)=R+\alpha R^{1/2}-\beta R^{-1/2}$  (see Fig.4b),
the curves start from the region $r>1,s<0$, cross the $\Lambda$CDM
line and tend to it. While for the model $f(R)=R+\alpha lnR+\beta$
(see Fig.4c), the curves generally do not cross the $\Lambda$CDM
line but approach the line in the future. When the parameters
$\alpha$ and $\beta$ have the same sign, the curves lie in the
region $r>1,s<0$, and in reverse the curves lie in the region
$r<1,s>0$ for the case that $\alpha$ and $\beta$ have the opposite
sign. From Fig.4, we can clearly find that the behavior of $r(z)$
and $s(z)$ for the models in the Palatini $f(R)$ theories is
different from $\Lambda$CDM model and even from other dark energy
models. Furthermore, various types of Palatini $f(R)$ theories can
be distinguished from each other in the $z-r$ and $z-s$ panels.

The evolutionary trajectories of the statefinder pairs $\{r,s\}$ and
$\{r,q\}$ can also help to differentiate various models in the
Palatini $f(R)$ theories from dark energy models. The trajectories
$r(s)$ and $r(q)$ for the models under consideration are plotted in
Fig.5. It is easy to see from the figure that various models in the
palatini $f(R)$ theories in the $s-r$ and $q-r$ planes exhibit the
significant differences. Furthermore, the evolutionary trajectories
$r(s)$ and $r(q)$ also effectively distinguish the $f(R)$ models
from the $\Lambda$CDM model and other dark energy models.

\begin{figure}[tbp]
\includegraphics[width=0.4\textwidth]{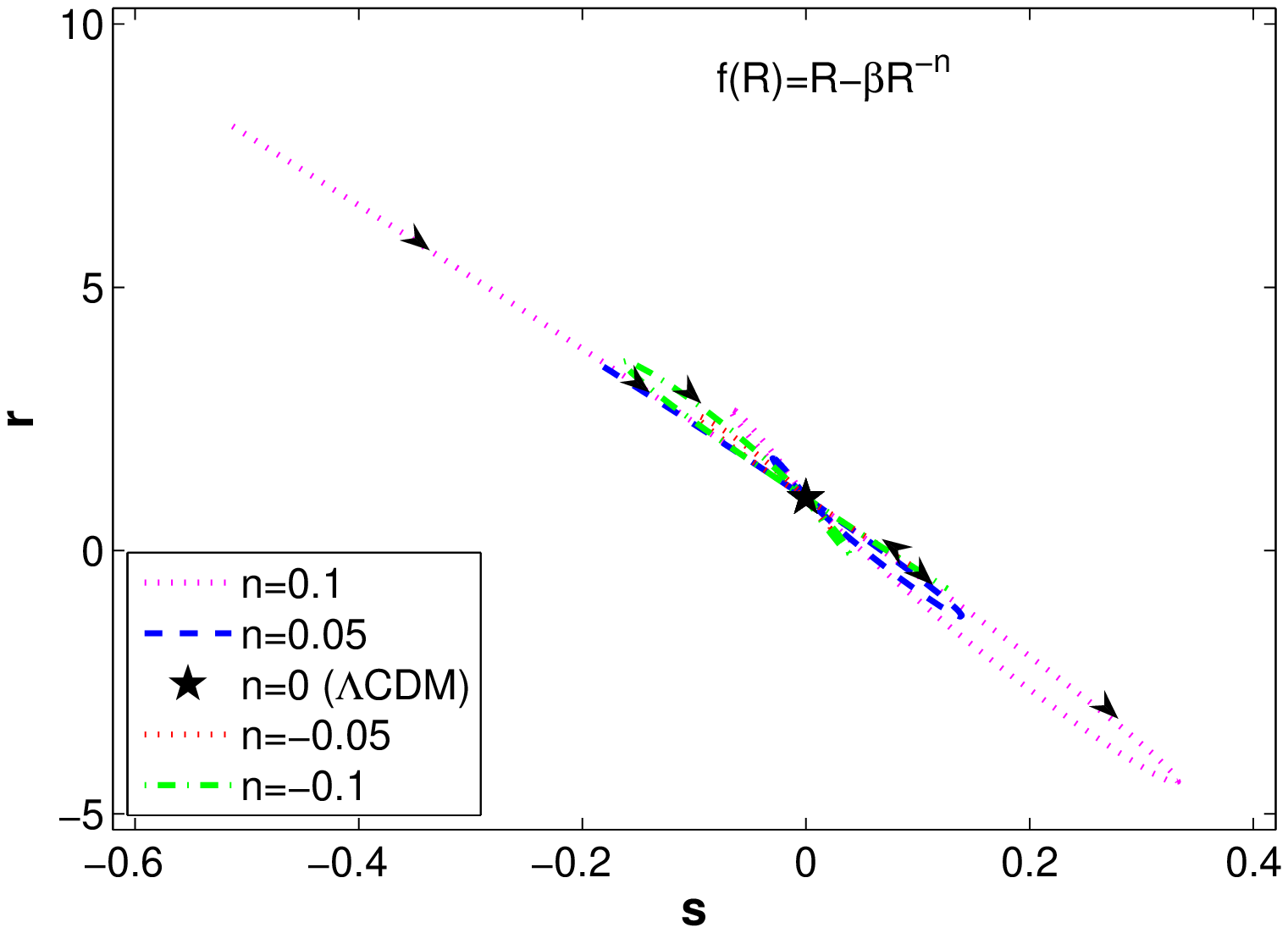}%
\includegraphics[width=0.4\textwidth]{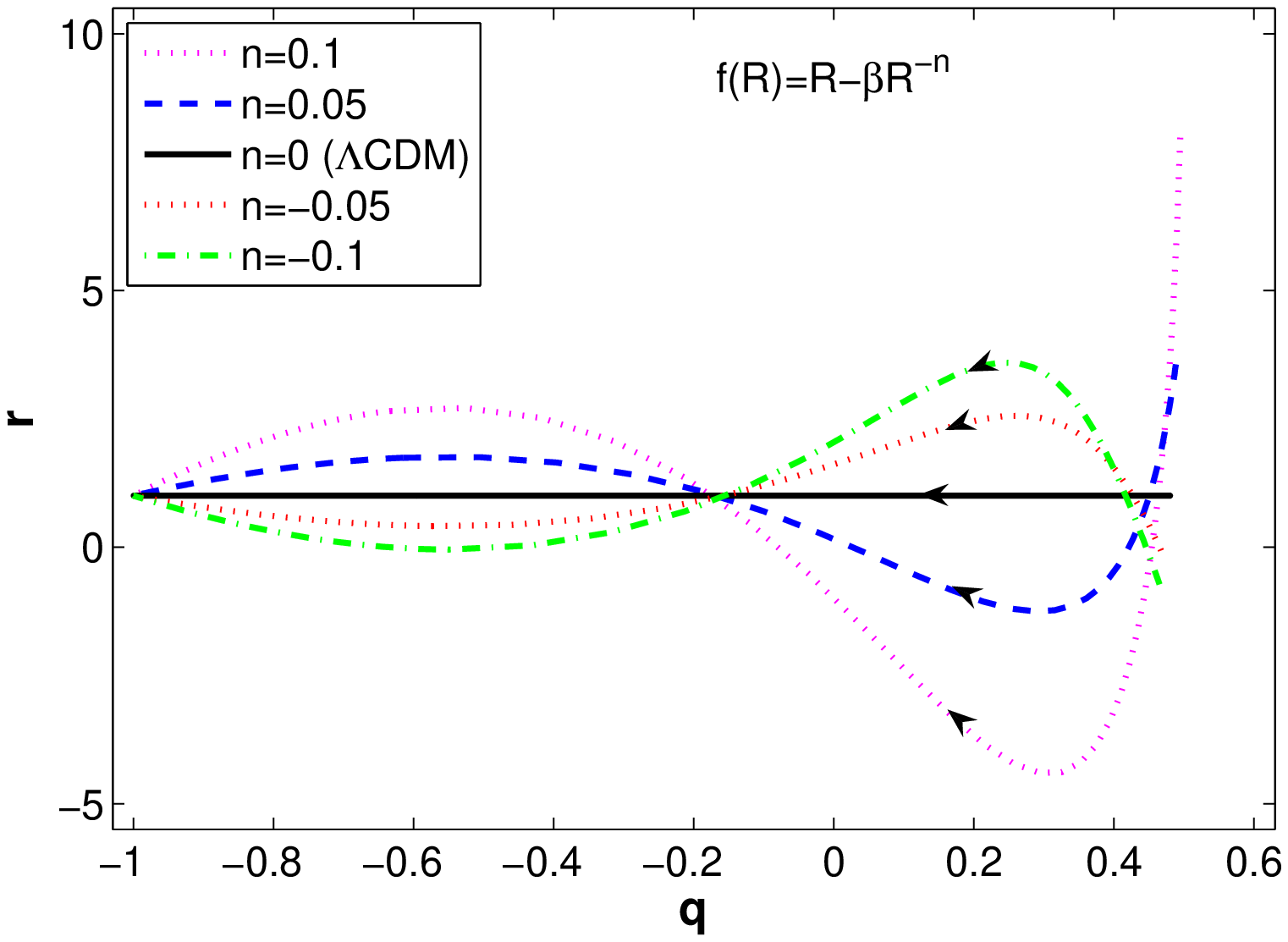}\\
\begin{center}
(a)
\end{center}
\includegraphics[width=0.4\textwidth]{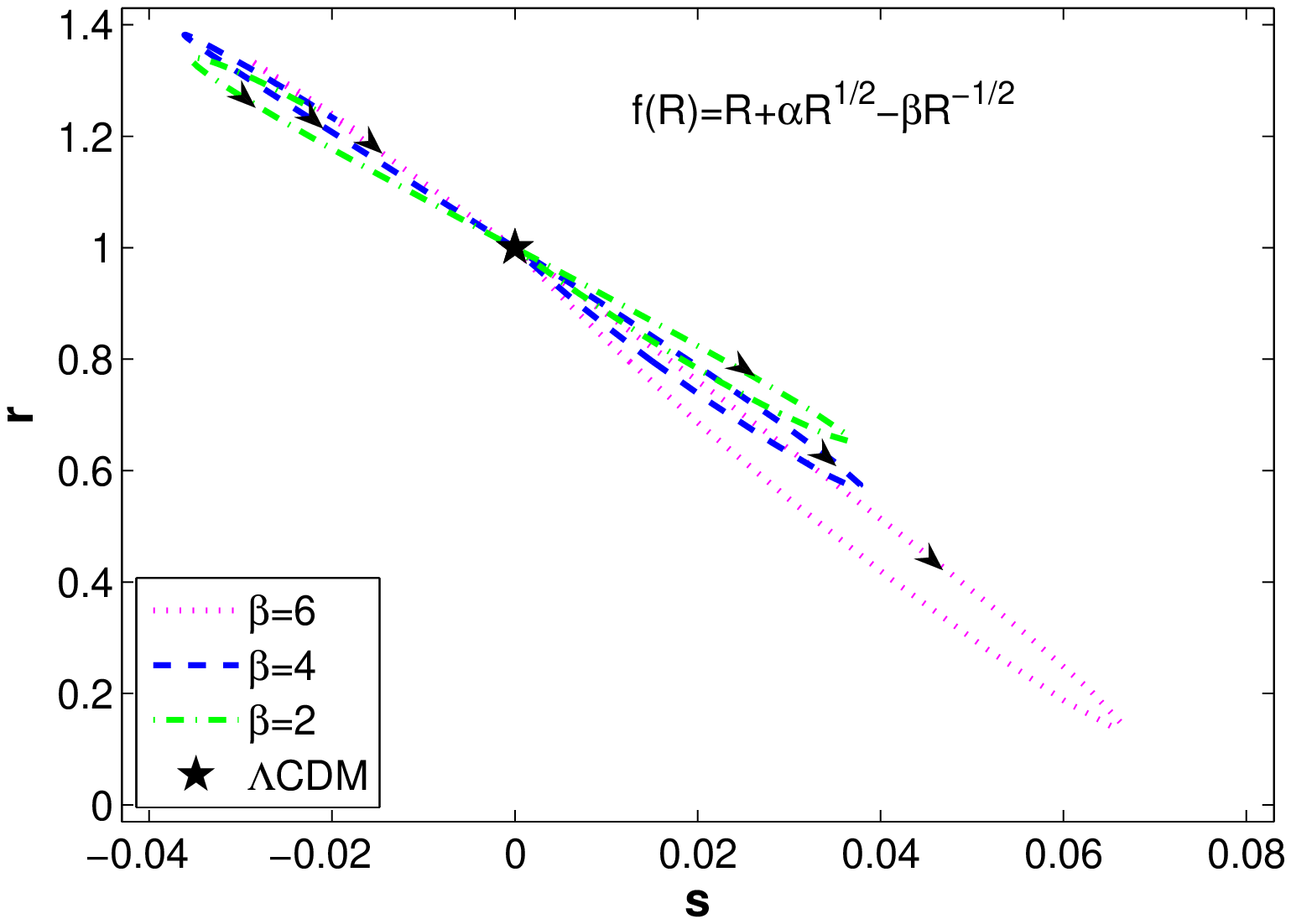}%
\includegraphics[width=0.4\textwidth]{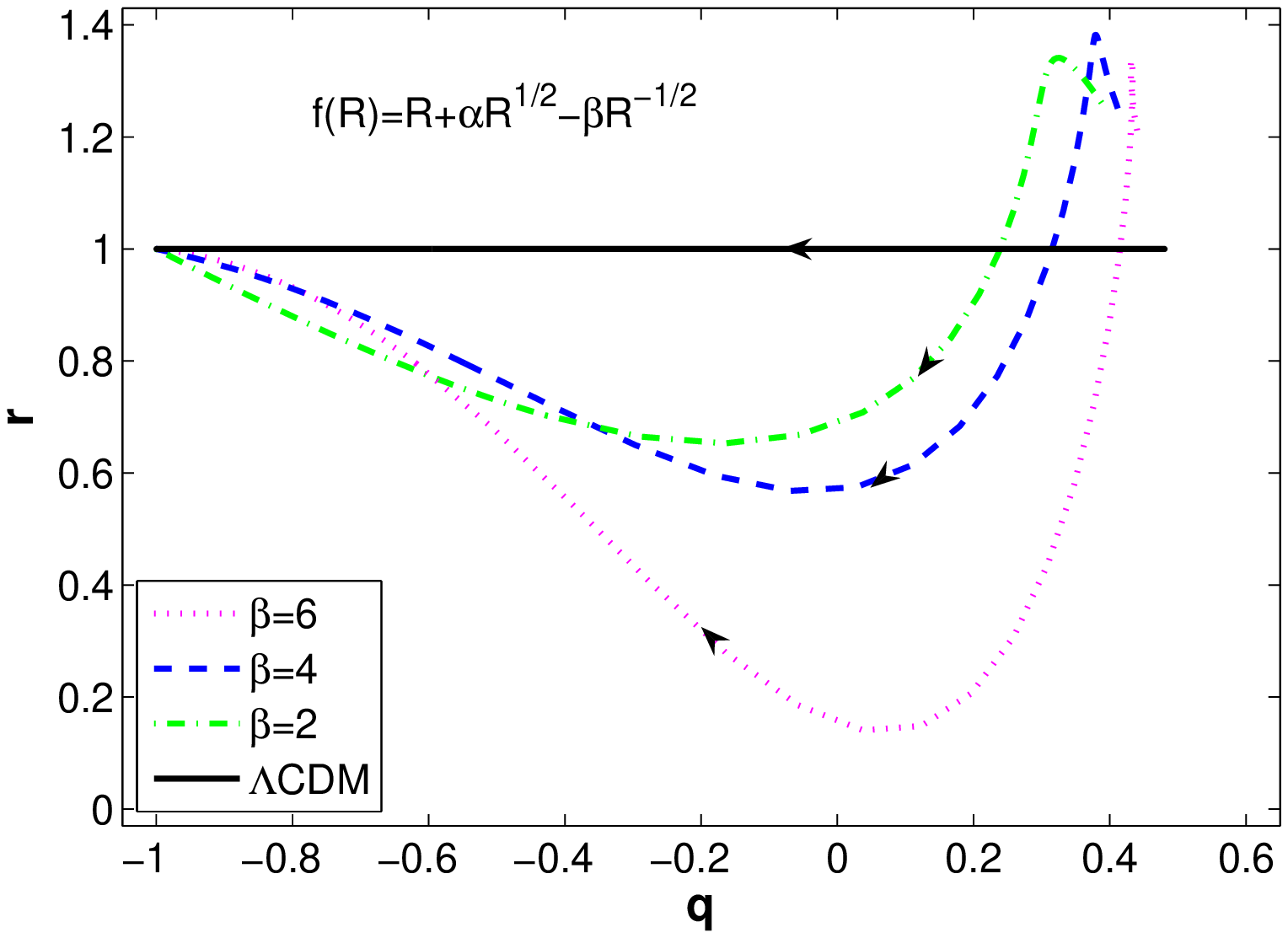}\\
\begin{center}
(b)
\end{center}
\includegraphics[width=0.4\textwidth]{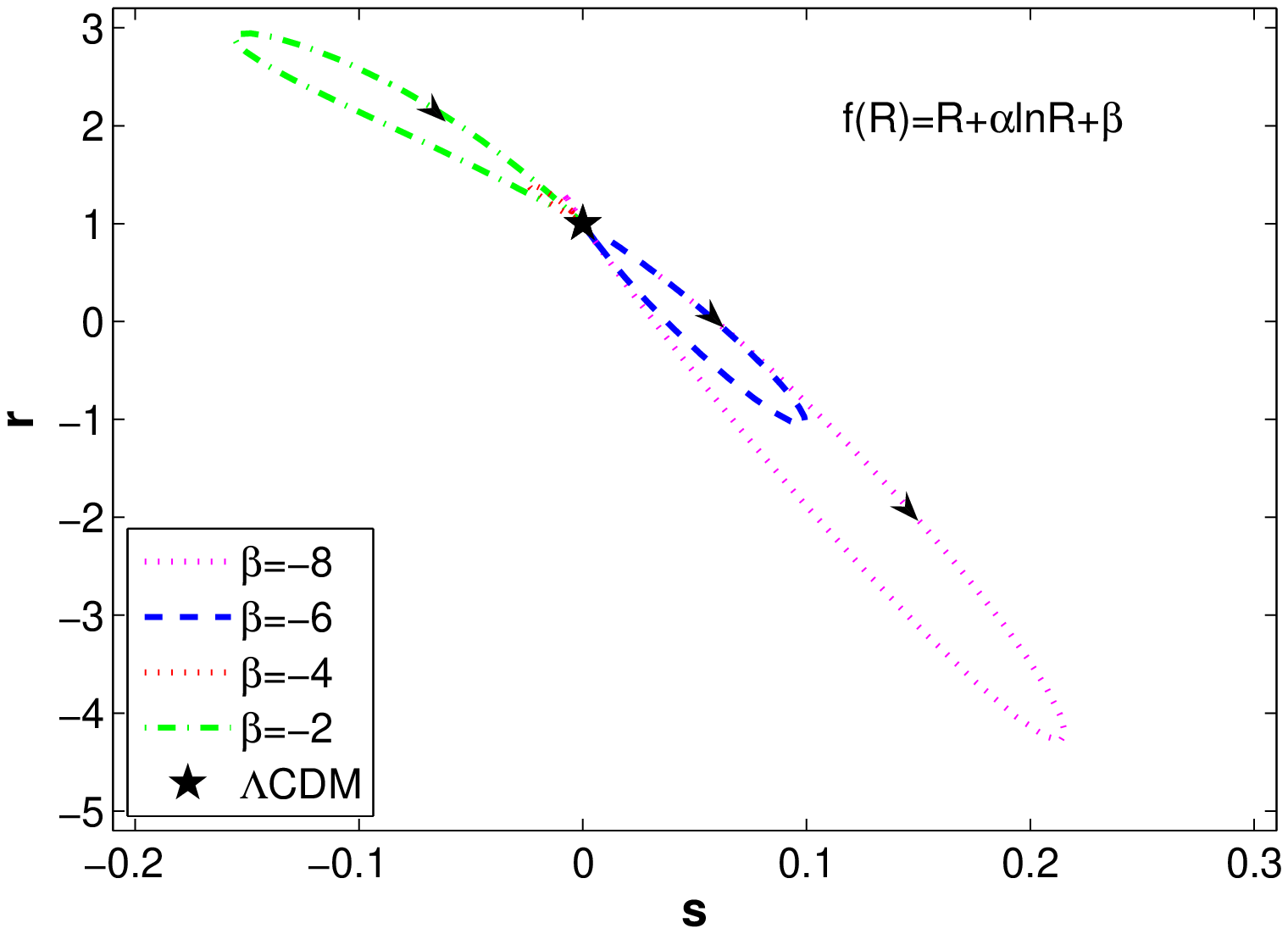}%
\includegraphics[width=0.4\textwidth]{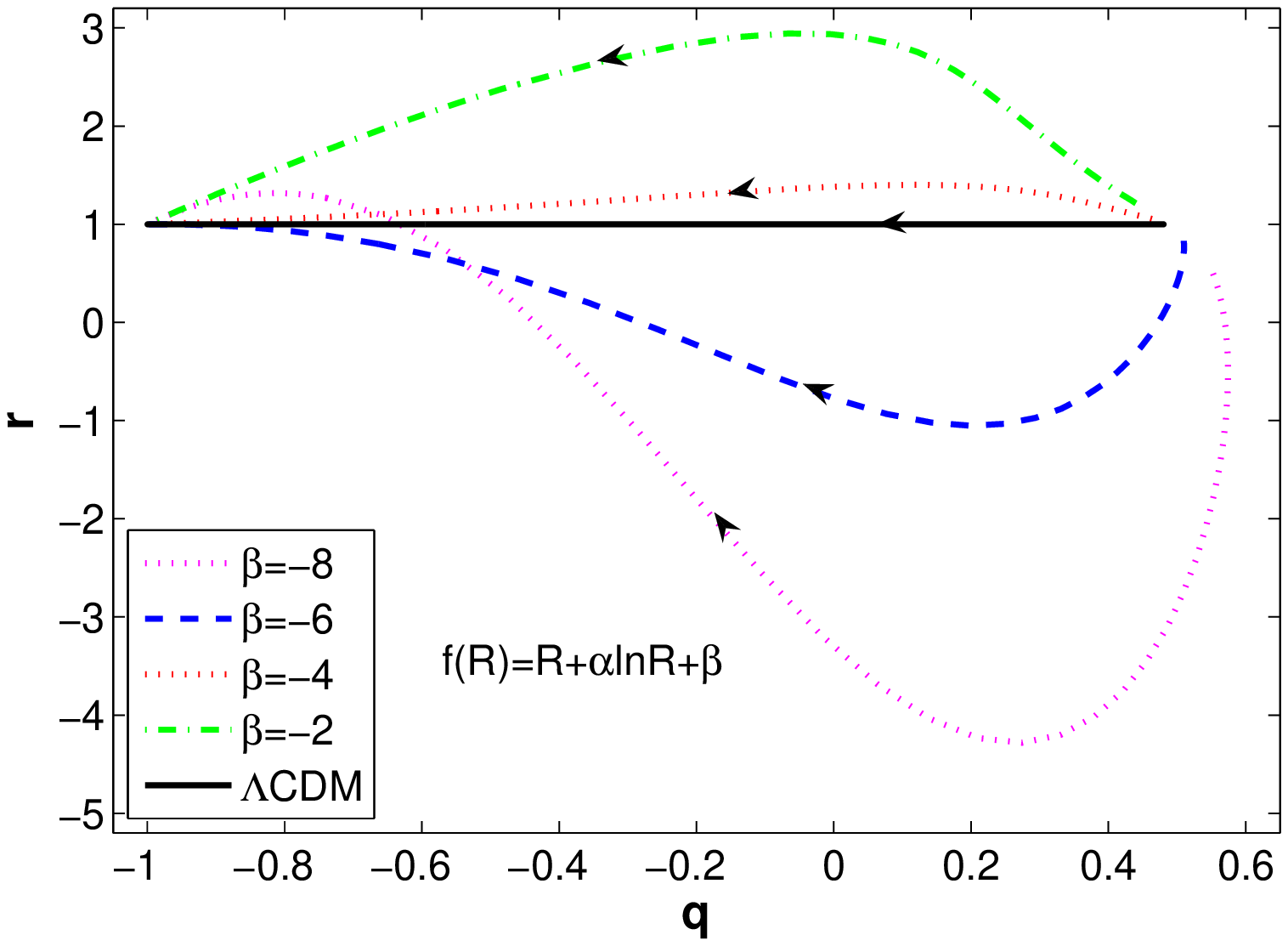}\\
\begin{center}
(c)
\end{center}
\caption{The evolution trajectories of $r(s)$ and $r(q)$ for the
models $f(R)=R-\beta R^{-n}$, $f(R)=R+\alpha R^{1/2}-\beta R^{-1/2}$
and $f(R)=R+\alpha lnR+\beta$ with $\Omega_{m0}=0.27$. }
\label{Fig.5}
\end{figure}

\section{Numerical Analysis from Observational Data}
In order to make $f(R)$ models to be compatible with cosmological
observations, we now turn to constrain the parameters in $f(R)$
models with the observational $H(z)$ data.

The Hubble parameter $H(z)$ data depends on the differential ages of
the universe as a function of redshift $z$ in the form
\begin{equation}
H(z)=-\frac{1}{1+z}\frac{dz}{dt},
\end{equation}
which provides a direct measurement for $H(z)$ through a
determination of $dz/dt$~\cite{price34}. By using the differential
ages of passively evolving galaxies determined from the Gemini Deep
Deep Survey (GDDS)~\cite{price35} and archival data~\cite{price36},
Simon et al. determined $H(z)$ in the range $0\leq z\leq1.8$ and
used them to constrain the dark energy potential and its redshift
dependence~\cite{price34}. In order to impose constraints on the
models of $f(R)$ gravity, we determine the best fit values for the
model parameters by minimizing
\begin{displaymath}
\chi^2=\sum_{i=1}^9=\frac{[H_{th}(z_{i}|s)-H_{obs}(z_{i})]^2}{\sigma^2(z_{i})},
\end{displaymath}
where $H_{th}(z_{i}|s)$ is the theoretical Hubble parameter at
redshift $z_{i}$ given by $(17)$; $H_{obs}(z_{i})$ are the values of
the Hubble parameter obtained from the data selected
by~\cite{price34}(SVJ05) and $\sigma(z_{i})$ is the uncertainty for
each of the nine determinations of $H(z)$.

\subsection{The Type $f(R)=R+\alpha R^{m}-\beta R^{-n}$}

\subsubsection{The Case of $f(R)=R-\beta R^{-n}$}

\subsubsection{The Case of $f(R)=R+\alpha R^{m}-\beta R^{-n}$ with $m=n=1/2$}

\subsection{The Type $f(R)=R+\alpha ln R+\beta$}

\section{Conclusions and Discussions}
In summary, we investigate the $f(R)$ gravity in Palatini formalism
by means of the statefinder diagnosis in this paper. Differences of
the evolutionary trajectories in the $s-r$ plane among a series of
$f(R)$ models have been found. Therefore, the statefinder parameters
are powerful to discriminate the models in the Palatini $f(R)$
gravity from dark energy models and even from other $f(R)$ models.
Also, the $q-r$ plane has been widely used for discussion on the
evolutionary property of the universe. We find that the $f(R)$
models under consideration exhibit different properties in the $q-r$
plane. Since the model parameters are found to be sensitive to the
$f(R)$ models, constraining the parameters in the models exactly
becomes a valuable task. We use the observational H(z) data to make
a combinational constraint.

\section*{Acknowledgements}
We are grateful to Xin-juan Zhang for helpful discussions. This work
was supported by the National Science Foundation of China (Grants
No.10473002, 10533010), 2009CB24901, the Scientific Research
Foundation for the Returned Overseas Chinese Scholars, State
Education Ministry.

\emph{Note added.}---As we were completing this paper we became
aware of the work reported in~\cite{price37}, which uses a similar
method to discriminate the $\Lambda$CDM from viable models of
$f(R)$. While the authors of that work consider the models which is
different from ours. Furthermore, we derive the statefinder
parameters directly from the Hubble parameter $H(z)$. In addition,
we impose constraints on the models by using different observational
data.

\end{document}